# Ferroelectric properties of SrRuO$_3$/BaTiO$_3$/SrRuO$_3$ ultrathin film capacitors free from passive layers

Y. S. Kim, J. Y. Jo, D. J. Kim, Y. J. Chang, J. H. Lee, and T. W. Noh

*ReCOE and School of Physics, Seoul National University, Seoul 151-747, Korea*

T. K. Song[a]

*Department of Ceramic Science and Engineering, Changwon National University, Changwon, Kyungnam 641-773, Korea*

J.-G. Yoon

*Department of Physics, University of Suwon, Suwon, Gyunggi-do 445-743, Korea*

J.-S. Chung

*Department of Physics, Soongsil University, Seoul 156-743, Korea*

S. I. Baik and Y.-W. Kim

*School of Materials Science and Engineering, Seoul National University, Seoul 151-744, Korea*

C. U. Chung

*Department of Physics, Hankuk University of Foreign Studies, Yongin, Gyunggi-do 449-791, Korea*

Structural studies on ultrathin SrRuO$_3$/BaTiO$_3$/SrRuO$_3$ capacitors, with BaTiO$_3$ thicknesses of between 5 nm and 30 nm, show well-defined interfaces between ferroelectric BaTiO$_3$ and electrode SrRuO$_3$ layers. In these capacitors, we cannot observe any extrinsic electrical effects due to either the formation of an insulating interfacial passive layer or passive-layer-induced charge injection. Such high quality interfaces result in very good fatigue endurance, even for the 5 nm thick BaTiO$_3$ capacitor.

---

[a] Electronic mail: tksong@changwon.ac.kr

Due to the current trend of ferroelectric (FE) device miniaturization, a lot of attention has been paid to ultrathin FE films.[1] As the film thickness decreases, interfaces between FE film and electrodes should play more important roles. Unfortunately for this development, numerous workers have reported the formation of a passive layer at the FE/electrode interface,[2-8] which results in the degradation of FE characteristics, such as a decrease in remnant polarization,[7] and an increase in the coercive field.[8] In addition, passive layers have been proposed to cause reliability problems, such as fatigue failures, in devices exploiting FE memory applications.[9] Especially, in ultrathin FE films, the existence of a passive layer could change FE characteristics drastically, by either affecting the formation of FE domains or inducing a very large depolarization field.[10]

In spite of its importance, it is not clear whether we will be able to fabricate an ideal FE capacitor free from a passive layer using an ultrathin FE film. At the FE/electrode interface, lattice mismatch, strain relaxation, charge trapping, interdiffusion, and/or defect clustering can all occur, resulting in the formation of an effective passive layer, whose properties are different from those of the FE film. Such passive layer formation is detrimental, especially for an ultrathin FE film. Up to this point, most analyses of ultrathin FE films (below 10 nm) rely on piezoelectric and/or structural measurements due to the large leakage current.[11,12] Recently, we reported on the fabrication of high quality $SrRuO_3$ (SRO)/$BaTiO_3$ (BTO)/SRO capacitors on $SrTiO_3$ (001) substrates, with thicknesses of between 5 and 30 nm.[13,14] From direct electrical measurements, we showed that their thickness dependent polarization scaling follows the prediction of the recent first principles calculation, given by Junquera and Ghosez.[15] In this Letter, we report further investigations of microstructure and FE characteristics of our capacitors, indicating the fact that even with a 5 nm BTO film, they should be free from a passive layer. Finally, we will show that they are also free from fatigue failure up to $10^9$ cycles.

High quality SRO/BTO/SRO heterostructures on SrTiO$_3$ (001) substrates were grown using pulsed laser deposition with reflection high-energy electron diffraction (RHEED). For clean interfaces, all of the fully strained BTO and the SRO layers were grown *in-situ* on TiO$_2$-terminated SrTiO$_3$ substrates under the same oxygen partial pressure and temperature.[16] Then, capacitors of 10×10 μm$^2$ size were fabricated by ion-milling. Details of the ultrathin film deposition and the capacitor fabrication were reported earlier.[13,14] Microstructure of the films was analyzed by transmission electron microscopy (TEM) and synchrotron x-ray reflectivity measurements. Polarization (*P*)-electric field (*E*) hysteresis loops were measured with a Sawyer-Tower circuit using a Yokogawa FG300 function generator and a DL7100 digital oscilloscope. Triangular waves of 20 kHz were used. Capacitance measurements were carried out with HP4294A Impedance Analyzer at 10 kHz by applying 10 kV/cm *ac*-field. Fatigue tests were performed with 100 kHz bi-polar pulses using an aixACCT TF Analyzer 2000.

Figure 1(a) shows a cross-sectional TEM image of 6.8 nm thick BTO capacitor. This image manifests atomically clean interfaces, marked by arrows. Since the TEM image can provide information only over a very small local area, we fabricated a BTO (6 nm)/SRO (21 nm) bi-layer and measured its x-ray reflectivity to check whether a clean interface can be formed over a much larger scale of about 100 μm. As shown in Fig. 1(b), x-ray reflectivity oscillations persist to a high angle, which corresponds to a wave vector *q* value of 4 nm$^{-1}$. Fig. 1(c) shows the Fourier-transform of the measured reflectivity, which represents the thickness profile of the bi-layer film. The sharp central peak near zero comes from a fade-out effect of the broad reflectivity envelope with an increase in the angle of incidence. The other three peaks at around 6, 21, and 27 nm are in good agreement with BTO, SRO, and BTO+SRO layer thicknesses monitored with RHEED.[17] These structural data clearly indicate that our capacitors could have very clean interfaces between the BTO and the SRO layers.

Electrical measurements have been known to be more sensitive to the passive layer formation than structural measurements. If there exists passive layers between the BTO and one of the SRO electrode layers (with an area of $A$), the series circuit model predicts that the capacitance ($C$) of the capacitor should have a relation to film thickness such that

$$A/C = t_F/\varepsilon_0\varepsilon_F + 2t_d/\varepsilon_0\varepsilon_d, \quad (1)$$

where $t_d$ ($t_F$) and $\varepsilon_d$ ($\varepsilon_F$) are the thickness and the dielectric constant of the passive (FE) layer.[5] We measured the $C$ values of our capacitors with and without a $dc$-bias field of 1300 kV/cm. As shown in Fig. 2, the plot of $A/C$ vs $t_F$ produces a linear curve, as expected from Eq. (1), with nearly zero intercepts with the $y$-axis. Some workers have argued that the finite screening length of electrons inside the electrode should result in a finite intercept.[18,19] We estimated such a contribution and found that it should be about 2 m$^2$/F, which is nearly comparable to the margin of error for our measurements.[20] The fact our $A/C$ vs $t_F$ curves have nearly zero intercepts within our experimental error range demonstrates absence of passive layers in our SRO/BTO/SRO capacitors.

During switching, the passive layer is usually exposed to a very high electric field and a charge injection through the passive layer can result in degradation of a device's performance, such as fatigue endurance.[9] Tagantsev and Stolichnov showed that the charge injection, approximated by the threshold conduction, will effectively screen the FE polarization by the charge of the accumulated carriers at the interfaces, and will change the detailed shape of the $P$-$E$ hysteresis loops.[2] The predicted dependence of the coercive field ($E_C$) and maximum polarization ($P_m$) is schematically shown in the inset of Fig. 3(c), and is found to be strongly dependent on the thickness of the ferroelectric layer.[2] Figure 3(a) and 3(b) shows hysteresis loops measured at various maximum applied fields for our BTO capacitors of 5 nm and 30 nm thicknesses, respectively. Amazingly, high quality $P$-$E$ hysteresis loops can be measured even for the 5 nm BTO capacitors. Note that our data show that $E_C$ increases with $P_m$. However, as shown in Fig. 3(c), the measured $E_C$ values do not show the predicted dependence on the BTO thickness: the $E_C$ values seem to be nearly independent of $t_F$. This result strongly suggests that our BTO capacitors are free from any charge injection effects due to passive layers with threshold conduction.

The high interface quality of our capacitors should provide good FE switching properties. Figure 4(a) shows that the remnant $P$ values of 5 nm and 30 nm thick BTO capacitors remain nearly constant after $10^9$ cycles of switching. In particular, the $P$-$E$ hysteresis loop of the 5 nm thick capacitor is almost the same after $10^9$ cycles as before cycling, as shown in Fig. 4(b). This fatigue-free characteristic might be originated from good interface properties of the FE/electrode interfaces.

In summary, structural and electrical analyses of our SrRuO$_3$/BaTiO$_3$/SrRuO$_3$ ultrathin film capacitors pointed out that passive layers are hardly observable. The ultrathin film capacitors showed good switching properties and were not degraded by fatigue stresses over $10^9$ cycles of switching. Our results showed the possibility of realizing ultrathin film electronic devices using ferroelectrics.

The authors acknowledge our valuable discussions with C. S. Hwang. This work was financially supported by the Korean Ministry of Education through the BK21 projects and by KOSEF through the CRI Program. The experiments at Pohang Light Source were supported by MOST and POSCO. T.K.S. acknowledges financial support in part by the Korean Council for University Education, for the 2004 Domestic Faculty Exchange.

FIG. 1. (a) TEM image of SRO/BTO (6.8 nm)/SRO capacitor structure. Arrows indicate the interfaces between BTO and SRO layers. (b) *q*-dependent x-ray reflectivity and (c) Fourier transform of (b) showing the thickness profile of BTO (6 nm)/SRO (21 nm) bi-layer.

FIG. 2. Reciprocal capacitances as functions of BTO thicknesses.

FIG. 3. *P-E* hysteresis loops for BTO capacitors of (a) 5 nm and (b) 30 nm thickness. (c) $P_m$ versus $E_C$ of BTO capacitors from 30 nm down to 5 nm thickness. Inset shows a schematic diagram according to the charge injection model,[2] where $E_{C0}$ is the $E_C$ value with the charge injection and $E_{th}$ is the threshold electric field for charge injection. Dotted, dashed, and dash dotted lines represent the simulation result of 5, 6.5, and 15 nm thick BTO capacitors, respectively.

FIG. 4. (a) Remnant *P* versus number of cycles for BTO capacitors of (solid) 5 nm and (open) 30 nm thickness. (b) *P-E* hysteresis loops for 5 nm thick BTO capacitor (circles) before and (squares) after $10^9$ cycles.

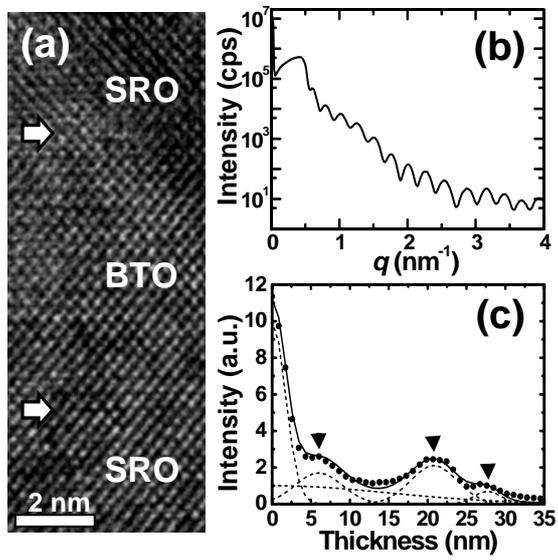

Fig. 1

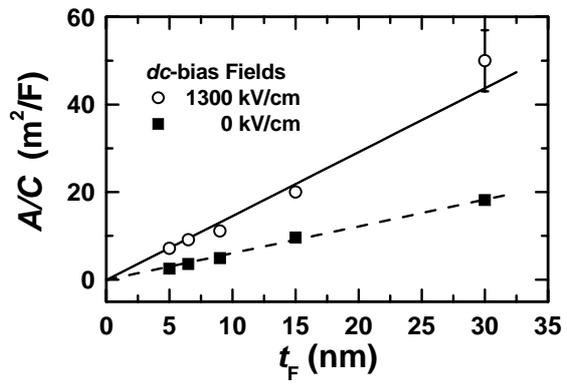

Fig. 2

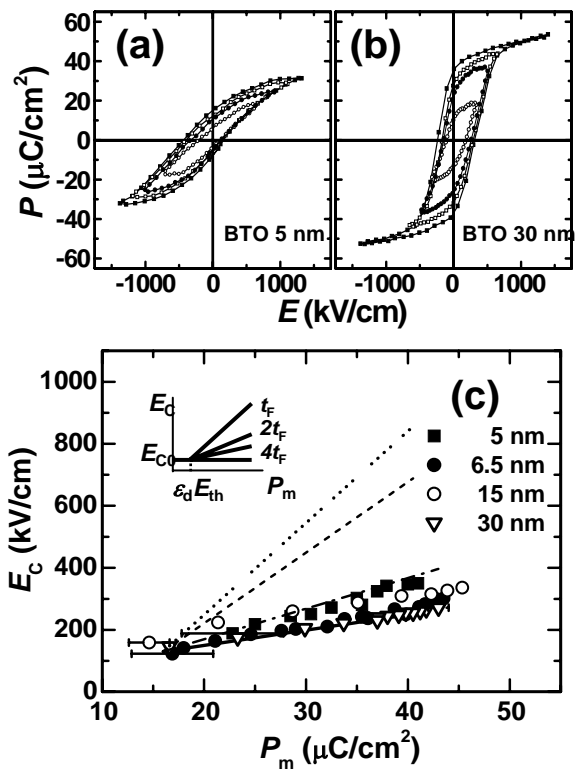

Fig. 3

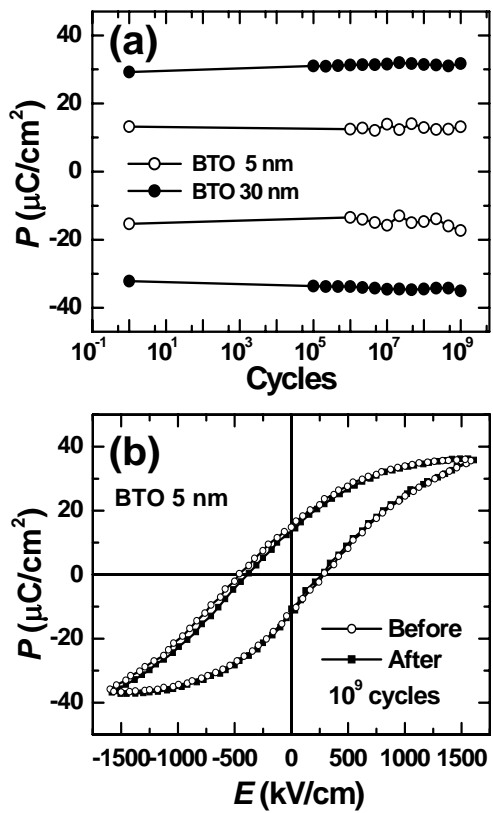

Fig. 4